\documentclass[useAMS]{mn2e}
\usepackage{psfig}
 \usepackage{url}
\usepackage{times}
\def\spose#1{\hbox to 0pt{#1\hss}}
\newcommand\lsim{\mathrel{\spose{\lower 3pt\hbox{$\mathchar"218$}}
     \raise 2.0pt\hbox{$\mathchar"13C$}}}
\newcommand\gsim{\mathrel{\spose{\lower 3pt\hbox{$\mathchar"218$}}
     \raise 2.0pt\hbox{$\mathchar"13E$}}}
\def\ltsima{$\; \buildrel < \over \sim \;$}
\def\lsim{\lower.5ex\hbox{\ltsima}}
\def\gtsima{$\; \buildrel > \over \sim \;$}
\def\gsim{\lower.5ex\hbox{\gtsima}}

\def\ergs{{\rm\thinspace erg \thinspace s^{-1}}}

\title[SDSS J102623.61+254259.5: a blazar at $z$=5.3]
{SDSS J102623.61+254259.5: the second most distant blazar at $z$=5.3}
\author[Sbarrato, T.]
{T.  Sbarrato$^{1,2},$\thanks{E--mail:
tullia.sbarrato@brera.inaf.it} G. Ghisellini$^2$, M. Nardini$^3$, G. Tagliaferri$^2$, L. Foschini$^2$,
G. Ghirlanda$^2$,
\newauthor{
F. Tavecchio$^2$, J. Greiner$^4$, A. Rau$^4$, N. Gehrels$^5$ }  \\
$^1$ Univ. dell'Insubria, Dipartimento di Fisica e Matematica, Via Valleggio 11, I--22100 Como, Italy \\
$^2$ INAF -- Osservatorio Astronomico di Brera, via E. Bianchi 46, I--23807 Merate, Italy \\
$^3$ Univ. di Milano Bicocca, Dip. di Fisica G. Occhialini, Piazza della Scienza 3, I--20126 Milano, Italy \\
$^4$ Max Planck Institut f\"ur extraterrestrische Physik, Giessenbachstrasse 1, 85748 Garching, Germany \\
$^5$ NASA--Goddard Space Flight Center, Greenbelt, Maryland 2077, USA
}

\begin{document}
%\date{Accepted 2011 January 06. Received 2011 January 06; in original
%  form 2010 November 16}

\pagerange{\pageref{firstpage}--\pageref{lastpage}} \pubyear{2012}

\maketitle
\label{firstpage}

\begin{abstract}
The radio--loud quasar SDSS J102623.61+254259.5, 
at a redshift $z$=5.3,
is one of the most distant radio--loud objects.
Since its radio flux exceeds 100 mJy at a few GHz, it is also one of the most powerful
radio--loud sources.
We  propose that this source is a blazar, i.e. we are seeing its jet at a small
viewing angle. 
This claim is based on the spectral energy distribution of this source,
and especially on its strong and hard X--ray spectrum, as seen by
{\it Swift}, very typical of powerful blazars.
Observations by the Gamma--Ray Burst Optical/Near--Infrared Detector (GROND)
and by the Wide--field Infrared Survey Explorer (WISE)  allow
to establish the thermal nature of the emission in the near IR--optical band.
Assuming that this is produced by a standard accretion disk, we derive that
it emits a luminosity of $L_{\rm d}\simeq9\times10^{46}$ erg s$^{-1}$
and that the black hole has a mass  between 2 and 5 billion solar masses.
This poses interesting constraints on the mass function of heavy ($>10^9 M_\odot$)
black holes at high redshifts.
\end{abstract}


%---------------------------------------------------------------------

\section{Introduction}
\label{sec-intro}

Blazars are Active Galactic Nuclei (AGN) whose relativistic jets are seen at 
a small angle from our line of sight, smaller than $1/\Gamma$, where
$\Gamma$ is the bulk Lorentz factor of the emitting plasma.
The radiation emitted by their jets is therefore strongly boosted,
making them visible at high redshifts.
The most distant blazar known is Q0906+6930 (Romani et al. 2004; Romani 2006) 
with a redshift $z=5.47$.
The spectral energy distribution (SED) of this source
reveals both the thermal (i.e. strong optical emission lines and continuum) and
boosted non--thermal (dominating in all bands but the near IR and optical) components.
This is a common characteristic of powerful blazars at high redshift: the non thermal
emission is characterized by two broad humps, peaking at lower 
frequencies as the bolometric luminosity increases, making the underlying thermal
emission visible in powerful blazars
(since it stands between the two non--thermal humps; Ghisellini et al. 2010a; 2010b).
The high energy hump, in these sources, peaks below 100 MeV, disfavoring its detection by the
Large Area Telescope (LAT) instrument onboard the {\it Fermi} satellite (Atwood et al. 2009). 
Instruments observing in the hard X--ray band can instead have more chances to detect
them, and in fact the Burst Alert Telescope (BAT) onboard the {\it Swift} satellite 
(Gehrels et al. 2004) have detected blazars up to larger redshifts than LAT 
(compare Ajello et al. 2009 and Ackermann et al. 2011).
Their X--ray spectrum is hard 
[i.e. $\alpha_X \lsim 0.5$, assuming $F(\nu) \propto \nu^{-\alpha_{x}}$], 
and this, together with a relatively strong X--ray to optical flux ratio, can be taken as a signature
of the blazar nature of the source.

In this letter we suggest that SDSS J102623.61+254259.5, 
a radio--loud AGN at $z=5.3$,
is a blazar, i.e. the viewing angle is smaller than $1/\Gamma$.
If so, then this source indicates the presence, at approximately the same redshift,
of  $\sim 2\Gamma^2$  other sources 
pointing in other directions, having similar 
intrinsic physical properties, including the black hole mass. 
Note that the flatness of the radio spectrum is not sufficient to guarantee the blazar nature 
of SDSS J102623.61+254259.5, since slightly misaligned sources (with respect to $1/\Gamma$)
still show a flat radio spectrum, especially at relatively large radio frequencies 
(at $z=5.3$, the observed 1.4 GHz flux corresponds to a rest frame frequency of $\sim$9 GHz), 
where the radio lobe steep radio emission does not contribute.
Evidences for its blazar nature are its very large radio--loudness 
% ($R\sim 5200$) 
and very large radio luminosity. 
Observed by {\it Swift}, it revealed a strong and hard X--ray flux,
confirming its blazar nature.
Observations performed by the Gamma--Ray Burst Optical/Near--Infrared Detector
(GROND; Greiner et al. 2008) and by 
the WISE satellite (Wright et al. 2010), together with the
Sloan Digital Sky Survey (SDSS; York et al. 2000) spectrum, allowed to constrain the
properties of the thermal emission.

% Because of the peculiar orientation of blazars, they are very important from the statistical 
% point of view in works that study the density of AGN in the early universe 
% (see as an example Volonteri et al., 2011).
% Indeed, for each blazar pointed toward us, we should expect to have a great number 
% ($N\sim 2 \Gamma^2 \sim 200$) of randomly oriented similar radio--loud AGN.

In this work, we adopt a flat cosmology with $H_0=70$ km s$^{-1}$ Mpc$^{-1}$ and
$\Omega_{\rm M}=0.3$.

%---------------------------------
\begin{table*} 
\centering
\begin{tabular}{llllllll}
\hline
\hline
~  &$g^\prime$ &$r^\prime$ &$i^\prime$ &$z^\prime$ &$J$ &$H$ &$K_{s}$ \\
\hline   
$\lambda_{\rm eff}$ (\AA)   &4587 &6220 &7641 &8999 &12399 &16468 &21706 \\  
AB magnitude      &$<$24.0 &22.07$\pm$0.08 &19.97$\pm$0.03 &19.86$\pm$0.04 &19.50$\pm$0.10 &19.23$\pm$0.15 &19.07$\pm$0.21 \\  
\hline
\hline 
\end{tabular}
\vskip 0.4 true cm
\caption{GROND AB observed magnitudes of B2 1023+25, taken 2012 April 16th 
(magnitudes not corrected for Galactic extinction). 
The first raw gives the effective wavelength of the filter.
}
\label{grond}
\end{table*}
%---------------------------------

\section{B2 1023+25 as a blazar candidate}
\label{sec-sample}

To obtain a sample of high redshift blazar candidates, we selected a list 
of quasars included in the SDSS DR7 Quasar catalog 
(Schneider et al. 2010), that have been analysed by Shen et al. (2011), 
with $z>4$ and radio--loudness $R>100$, defined as $R=F({\rm 5GHz})/F({\rm 2500\AA})$ (rest frame)
in the work by Shen et al. 
We required a high radio--loudness in order to preferentially select
objects with the jet directed toward us.
Through this selection, we obtained a sample of 31 sources with $z>4$
including 3 with $z>5$. 
The latter sources have no rest frame 2500 \AA\ flux and hence 
we calculated $F({\rm 2500\AA})$ extrapolating from the continuum at $\lambda=1350\rm\AA$.

SDSS J102623.61+254259.5 (= B2 1023+25) 
is the best blazar candidate from this list, because of its 
 very high redshift and 
extreme radio--loudness ($R\sim 5200$).
The precise value of the redshift
is somewhat uncertain, since $z=5.284$ in NED, taken from the 
6th Data Release of the Sloan survey (DR6), while $z=5.3035$
in DR7, and $z=5.266$ in DR8.
In this paper we assumed therefore a value of $z$=5.3.
This is a known quasar, observed for the first time in 
the B2 Catalog of radio sources (Colla et al. 1972).
The radio position of this source is RA=10h26m23.62s, DEC=+25d42m59.4s and, 
according to the CRATES All--Sky Survey (Healey et al. 2007), the source has a flat spectrum 
with $\alpha_{1.4/4.8}=0.489$ [between 1.4 and 4.8 GHz; $F(\nu)\propto \nu^{-\alpha}$]
and $\alpha_{1.4/8.4}=0.504$ with a radio flux of $F_{1.4{\rm GHz}}=260.7{\rm mJy}$ and
$F_{8.4{\rm GHz}}=105.7{\rm\ mJy}$.
% $F_{4.8{\rm GHz}}=142\ {\rm mJy}$. 
% $F_{1.4{\rm GHz}}=260.7{\rm mJy}$
We retrieved a number of radio fluxes from archival data\footnote{
ASI Science Data Center (ASDC): \url{http://tools.asdc.asi.it/}
}.
Moreover, B2 1023+25 is included in the WISE All--Sky Source Catalog\footnote{
Data retrieved from \url{http://irsa.ipac.caltech.edu/}
}, with clear detections in the two bands at lower wavelengths of the instrument, 
i.e.\ $\lambda=3.4\mu$m and $\lambda=4.6\mu$m.

We retrieved and analyzed data of {\it Fermi}/LAT (Atwood et al., 2009) collected 
in the period 2008 August 4th -- 2012 June 14th ($\sim 3.8$ yrs). 
Photons of class 2 (``source"), with energies in the range 0.1--100 GeV, 
collected in a region of $10^{\circ}$ radius from the source position, 
were analyzed by using LAT Science Tools v. 9.27.1, Instrument Response Function P7, 
and the corresponding background maps. 
The followed procedures are described in detail in the analysis threads of the 
{\it Fermi} Science Support Center\footnote{http://fermi.gsfc.nasa.gov/ssc/data/analysis/}. 
Particularly, the sources B2 1023+25 and Q0906+693 were fitted with a power law model, 
together with any other source of the 2FGL catalog (Nolan et al., 2012) 
within $10^{\circ}$ from their sky coordinates. 
The analysis resulted in no detection for both sources, 
with a $5\sigma$ upper limit on the photon flux above 100 MeV
of $3.0\times 10^{-9}$~ph~cm$^{-2}$~s$^{-1}$.

\subsection{GROND observations}

The 7--band  GROND imager, mounted at the 2.2 m MPG/ESO telescope at La Silla Observatory
(Chile), started observing B2 1023+25 on 2012, April 16th at 23:16:56 UTC. 
We carried out two 8--minutes observations simultaneously in all 
7 $g^\prime, r^\prime, i^\prime, z^\prime, J, H, K_{\rm s}$ bands for a total exposure 
time of 919 s in the optical as well as 960 s in the NIR bands. 
Observations were carried out at an average seeing of 1.2$\arcsec$ and at an average airmass of 2.1. 
The source was  clearly detected in all bands but $g^\prime$ for which an upper limit of 
24.0 (AB magnitude) can be given.

The GROND  optical and NIR image reduction and photometry were
performed using standard IRAF tasks (Tody 1993) similar to the
procedure described  in Kr\"uhler et al. (2008). 
A general model for the point--spread function (PSF) of each image was
constructed using bright field stars, and it was then fitted to the point source. 
The absolute calibration of the $g^\prime, r^\prime, i^\prime, z^\prime$ bands was 
obtained with respect to the magnitudes of SDSS stars within the 
blazar field while the $J, H, K_{\rm s}$ 
bands calibration was obtained with respect to magnitudes of 
the Two Micron All Sky Survey (2MASS) stars (Skrutskie et al. 2006).

Tab. \ref{grond} reports the observed AB magnitudes, not corrected for 
the Galactic extinction of $E(B-V)=0.02$ from Schlegel et al. (1998).

\subsection{{\it Swift} observations}

Since this is our best blazar candidate, we 
requested a ToO observation from the {\it Swift} 
team, in order to have X--ray data to confirm our hypothesis.
The {\it Swift} satellite observed B2 1023$+$25 twice: on 2012, June 21st (ObsID: 00032500001) 
and on 2012, June 22nd (ObsID: 00032500002). 
Data of the X--ray Telescope (XRT, Burrows et al. 2005) and the UltraViolet Optical Telescope 
(UVOT, Roming et al. 2005) were downloaded from HEASARC public archive, 
processed with the specific {\it Swift} software included in the package 
{\tt HEASoft v. 6.12}\footnote{
Including the XRT Data Analysis Software (XRTDAS) developed under 
the responsibility of the ASI Science Data Center (ASDC), Italy.
}, and analysed. 
The calibration database was updated on July 16, 2012. 
We did not consider the data of the Burst Alert Telescope (BAT, Barthelmy et al. 2005), 
given the weak X--ray flux.

The total exposure on the XRT was of $\sim$10 ks and resulted in 26 counts. 
Given the low statistics, the fit with a power law model with Galactic absorption 
($N_{\rm H}=1.50\times 10^{20}$~cm$^{-2}$, Kalberla et al. 2005) 
was done by using the likelihood (Cash 1979). 
The output parameters of the model were $\Gamma = 1.1\pm 0.5$ 
and an integrated observed flux 
$F_{0.3-10\,\rm keV}=(1.7\pm 0.4)\times 10^{-13}$~erg~cm$^{-2}$~s$^{-1}$. 
The X--ray data displayed in the SED (Fig. \ref{sed}) has been rebinned 
to have $3\sigma$ in each bin.

UVOT observed the source with two filters only: $U$, with 8.4 ks exposure, 
and $UVW1$, with 1.6 ks. 
Both resulted in no detection with $3\sigma$ upper limits of the observed magnitudes 
equal to 21.70 and 20.76 mag, respectively.

\subsection{Estimate of the disc luminosity}
\label{Ld}

We can calculate the disc luminosity $L_{\rm d}$ in two ways.
The first is by assuming that the IR--optical flux
is produced by a standard, Shakura \& Sunjaev (1973) accretion disc,
and finding the best $L_{\rm d}$ that accounts for the data.
In this respect, for B2 1023+25 we have a rather good set of IR--optical data 
showing that the peak of the disc emission is at frequencies just 
below the absorption caused by intervening clouds (see Fig. \ref{mass}).
Since the overall $L_{\rm d}$ is about twice the $\nu L_\nu$ peak luminosity,
we directly derive $L_{\rm d}= \simeq9\times10^{46}\ergs$.

The second method is through the broad emission lines, 
assuming they re--emit a fraction $C$ (i.e. their covering factor) of the
ionizing luminosity, and assuming that the latter is provided by
the accretion disc.
The average value of $C$ is around 0.1 (Baldwin \& Netzer, 1978; Smith et al., 1981),
with a rather large dispersion.
B2 1023+25 shows a prominent broad Ly$\alpha$, that unfortunately is absorbed in its blue wing.
Using only the red wing, and subtracting the continuum in the same frequency
range (adopting a power law model), we derive 
$L_{\rm Ly\alpha}\sim 2\times10^{45}\ergs$ and a FWHM=4000 km s$^{-1}$.
These are twice the values we measured for the red wing
only, implicitly assuming that the line is intrinsically symmetric.
The ratio between the entire BLR luminosity and $L_{\rm Ly\alpha}$
has been calculated by Francis et al. (1991) and  Vanden Berk et al. (2001).
The former gives a $L_{\rm BLR}/L_{\rm Ly\alpha}=5.55$, while the latter gives a 
$L_{\rm BLR}/L_{\rm Ly\alpha}=2.7$.
Therefore $L_{\rm BLR}\sim$(5--10)$\times 10^{45}$ erg s$^{-1}$.
Finally, assuming $C=0.1$, the estimate for the disc luminosity is 
$L_{\rm d}\sim$(5--10)$\times 10^{46}$ erg s$^{-1}$.
Although this estimate is affected by relevant uncertainties 
(e.g. partially absorbed Ly$\alpha$ line and uncertain value of $C$),
the estimated range encompasses the value found with the first method.

\section{Black Hole Mass estimate}
\label{mass}

The flux and spectrum of the disc depend on the central
black hole mass $M_{\rm BH}$ and on the accretion rate $\dot{M}$.
The latter is traced by the total disc luminosity $L_{\rm d}=\eta\dot{M}c^2$,
with the efficiency parameter $\eta$ assumed to be $\eta=0.08$.
If we measure $L_{\rm d}$ (hence $\dot M$), we are left with $M_{\rm BH}$
as the only remaining free parameter. For a given $\dot M$, the black hole mass
determines the peak frequency of the disc emission.
A larger mass implies a larger disc surface and hence a lower temperature 
emitting a given luminosity. 
Therefore, for a fixed $L_{\rm d}$, a larger $M_{\rm BH}$ shifts the peak 
to lower frequencies. 
Then the best agreement with the data fixes $M_{\rm BH}$.
We adopt $L_{\rm d}=9\times 10^{46}$ erg s$^{-1}$, as discussed above,
and calculate the overall spectrum from a standard Shakura \& Sunjaev (1973) disc.

% Unfortunately, we have some uncertainties in this process, since we have a range of possible
% values of $L_{\rm d}$ derived from 
% $L_{\rm BLR}$ (due to the different templates and to the uncertain value of $C$).
% On the other hand, the rather good set of IR--optical data shows that the peak of the 
% disc emission is at frequencies just below the absorption caused by intervening clouds 
% (see Fig.\ \ref{mass}), 
% and has a luminosity $\nu L_{\rm \nu,peak}\simeq 4.5\times10^{46}\ergs$, corresponding to
% an overall disc luminosity $L_{\rm d} \simeq9\times10^{46}\ergs$.
% Comparing it with $L_{\rm BLR}$ derived above, we obtain $C\sim$5--10\%,
% in good agreement with standard assumptions.
% This leaves $M_{\rm BH}$ as the only free parameter, that we find by
% fitting the data with our disk model.
Fig. \ref{mass} shows the data in the IR--optical band (as labelled) and three 
disc emission spectra calculated assuming the same $L_{\rm d}$ and 
three different mass values:
$M_{\rm BH}=4.5\times10^9M_\odot$ (green line), 
$2.8\times10^9M_\odot$ (blue) and $1.8\times10^9M_\odot$ (cyan).
Fig. \ref{mass} shows that, outside this range of masses, 
the model does not satisfactory reproduce the data any longer.

%--------------------------------------------------
\begin{figure}
\hskip -0.3cm
\psfig{file=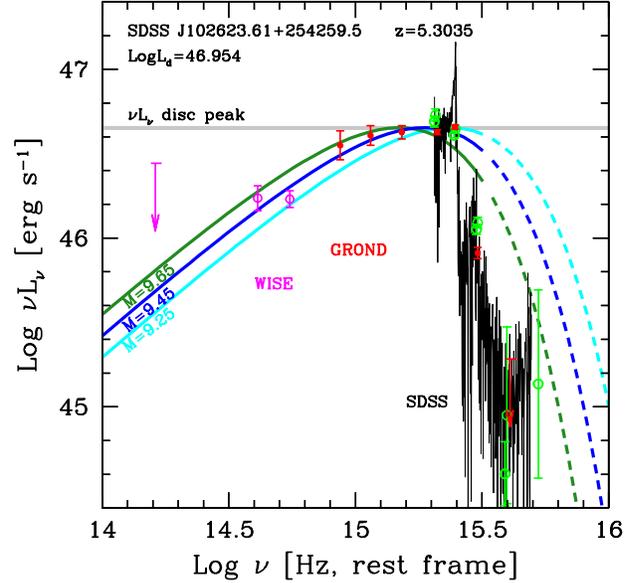,height=9.2cm,width=9.2cm}
\vskip -0.5 cm
\caption{
Optical--UV SED of B2 1023+25 in the rest frame, together with 
models of standard accretion disc emission.
Data from WISE and GROND, and the SDSS spectrum are labelled.
Green points are archival data taken from ASDC SED builder.
The grey stripe indicates the $\nu L_\nu$ peak luminosity of the disc.
We show the spectrum of three accretion disc models
with the same luminosity and different $M_{\rm BH}$:
$\log(M_{\rm BH}/M_\odot)$=9.65 (green), 9.45 (blue) and 9.25 (cyan).
Note that outside this range of masses, the model cannot fit 
satisfactory the data.
}
\label{mass}
\end{figure}
%--------------------------------------------------

%--------------------------------------------------
\begin{figure}
\hskip -0.3cm
\psfig{file=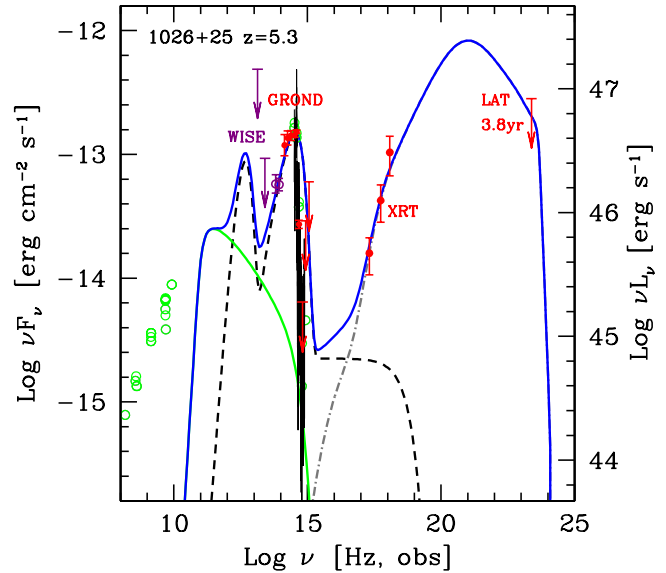,height=9.cm,width=9.cm}
\vskip -0.5 cm
\caption{
SED of B2 1023+25 together with the adopted model. 
Data from WISE, GROND, {\it Swift}/XRT and {\it Fermi}/LAT
are labelled.
Green points are archival data taken from ASDC SED builder.
The solid green line is the synchrotron component of the model,
the dashed black line is the torus+disc+X--ray corona 
component, the dot--dashed grey line is the External Compton
contribution.
The LAT upper limits is for 3.8 years, 5$\sigma$.
} 
\label{sed}
\end{figure}
%--------------------------------------------------

%--------------------------------------------------
\begin{figure}
\hskip -0.3cm
\psfig{file=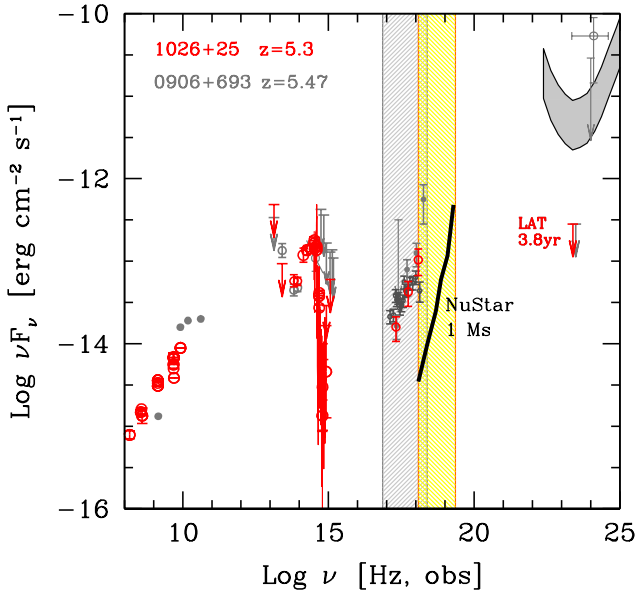,height=9cm,width=9cm}
\vskip -0.5 cm
\caption{
Comparison between the SED of B2 1023+25 (red symbols) and of Q0906+6930
(grey symbols).
As can be seen, the two sources are very similar.
The grey vertical stripe marks the [0.3--10 keV] energy range of XRT,
while the yellow vertical stripe corresponds to the [5--80 keV] energy range of the {\it nuStar}
satellite. 
The curved grey stripe corresponds to the sensitivity of {\it Fermi}LAT after 1 yrs of
operation (5$\sigma$, lower bound) and of 3 months (10 $\sigma$, upper bounds).
We show the sensitivity ($3\sigma$) of the {\it NuStar} satellite for an exposure of 1 Ms.
} 
\label{comparison}
\end{figure}
%--------------------------------------------------

\section{Overall spectral energy distribution}

Fig. \ref{sed} shows the overall SED of B2 1023+25, together with a fitting model.
The latter is described in Ghisellini \& Tavecchio (2009), 
and consists of one emitting zone, in which relativistic electrons,
whose energy distribution is derived through a continuity equation,
emit by the synchrotron and Inverse Compton processes.
The accretion disk component is accounted for, as well the
infrared emission reprocessed by a dusty torus and the X--ray 
emission produced by a hot thermal corona sandwiching the accretion disc.
The parameters adopted to fit the SED are very similar to other powerful blazars
studied and interpreted with the same model (Ghisellini et al., 2010a; 2010b).
The bulk Lorentz factor is $\Gamma=14$, the size of the emitting region is
$R=5\times 10^{16}$ cm, located at a distance of $5\times 10^{17}$ cm from
a black hole of $M_{\rm BH}=2.8\times 10^9 M_\odot$.
The magnetic field is $B=2.1$ G.
The main emission process for the X--ray and $\gamma$--ray non--thermal continuum
is the inverse Compton process off the emission line photons (i.e. External Compton).
Since the emitting region is rather compact, its radio emission is self--absorbed
(up to$\sim$600 GHz) and cannot account for the observed radio flux, that must necessarily
come from much larger zones.
The Poynting flux carried by the jet is $P_{\rm B} \sim 10^{46}$ erg s$^{-1}$,
and the power spent to produce the radiation we see is $P_{r}\sim 3\times 10^{45}$ erg s$^{-1}$.
The power in the bulk motion of the emitting relativistic electrons is $P_{\rm e}\sim 2\times 10^{44}$
erg s$^{-1}$, while the power in cold protons (assuming one proton per relativistic electron)
is $P_{\rm p}\sim 5.5\times 10^{46}$ erg s$^{-1}$.

\subsection{Comparison with Q0906+6930}

Fig. \ref{comparison} shows how the SED of B2 1023+25 compares with the SED
of Q0906+693.
As can be seen, the two SEDs are very similar.
The only, remarkable, difference could be in the $\gamma$--ray band, if
the tentative EGRET association with Q0906+6930 is real.
On the other hand, the upper limits of {\it Fermi}/LAT
are the same for the two sources.
This may suggest that the EGRET flux is not produced by Q0906+6930, although
variations of even more than two orders of magnitude of the $\gamma$--ray
flux are not uncommon for blazars (see e.g. Abdo et al. 2011; Ghirlanda et al. 2011).

Fig. \ref{comparison} shows, as vertical stripes, the energy bands of XRT and
of {\it NuStar} (Harrison et al., 2010).
We report also the pre--launch estimated {\it NuStar} sensitivity for an exposure of 1 Ms.
It is remarkable that both sources have an X--ray flux lying
one order of magnitude above the {\it NuStar} sensitivity,
making their detection possible even with a moderate exposure time,
and even at the high energy limit of the instrument, i.e. at $\sim$80 keV.
Note that 80 keV correspond, in the rest frame of a $z=5.3$ source,
to $\sim$500 keV, that could be rather close to the peak of the high energy emission.

\section{Discussion and conclusions}
\label{sec-discussion}

In this letter we propose that the radio--loud, high redshift quasar B2 1023+25 
is a blazar.
It is the second most distant blazar known up to now, with redshift 
$z=5.3$, that corresponds to an age of the Universe of $10^9$yrs.

Both its thermal and non--thermal components are very luminous. 
In agreement with the blazar sequence (Fossati et al. 1998) the two broad 
non--thermal humps peak at small frequency.
In particular, the hard X--ray spectrum and the upper limit in the $\gamma$--ray band 
constrain the high energy component to peak in the MeV region of the spectrum.
Therefore this source, along with the similar other powerful blazars,
should have a relatively large hard X--ray flux and would have been an
ideal target for planned hard X--ray instruments, such as 
{\it EXIST}, {\it Symbol--X} and {\it NHXM},
and it is a good target to be observed with the 
only orbiting satellite with focussing hard X--ray telescopes, i.e. 
{\it NuStar} (Harrison et al., 2010). 
The great sensitivity over its energy range [5--80 keV] enables it to 
detect the hard X--ray spectrum of this source, even if it cannot 
directly observe the peak of the high energy hump.

\vspace{0.3 cm}

Thanks to the good data coverage in the IR--optical range, 
given by the combination of the 7 simultaneous optical--near IR 
filters of GROND coupled with WISE and SDSS data, 
we were able to estimate the central black hole mass, 
that turned out to be between 2 and 5 billion solar masses.
This relatively small range
compares well with the virial estimates in general
(that have at least a factor $\sim$4 of uncertainties, see e.g.
Vestergaard \& Peterson 2006), and, in particular,
it is the only method applicable for this source, where only a partially
absorbed Ly$\alpha$ line is visible (and virial methods are not yet well 
calibrated in this case).
Near--IR spectroscopy, and hence the detection of the broad
CIV and MgII lines, would help to improve the virial estimate
of the black hole mass and be a valuable cross--check 
of our estimate, based on the disk--continuum.

It is remarkable to find radio--loud AGN hosting BH that massive 
at an age of the Universe of only $10^9$yrs.
The discovery that B2 1023+25 is a blazar, together with that one of Q0906+693, 
in fact, implies the existence of an entire 
family of radio--loud, very massive quasars in the early Universe, whose
jet is not aligned with our line--of--sight.
Moreover, it is possible to roughly estimate the number of these objects and their
spatial density.

The comoving density of heavy black holes at high redshifts
of radio--loud sources has been studied by Volonteri et al. (2011),
based on the 3 years BAT catalog and the blazar luminosity function,
in hard X--rays, derived by Ajello et al. (2009), and modified
(beyond $z=4.3$) by Ghisellini et al. (2010a).
In the latter paper the observational constrain on the blazar density
with black holes heavier than $10^9 M_\odot$ in the redshift bin
$5<z<6$ was based on the detection of only one object: Q0906+6930.
Assuming it was the only blazar in the entire sky in this
redshift bin, Ghisellini et al. (2010a) derived a comoving density of 
$2.63\times 10^{-3}$ Gpc$^{-3}$ of blazars hosting an heavy black hole
in the $5<z<6$ bin (see Fig. 15 in that paper).

B2 1023+25 was selected in the SDSS catalog covered by FIRST observations,
and the common area of the sky of these two surveys is 8770 square degrees.
Its existence in this portion of the sky implies
that the space density of blazars
hosting a $M_{\rm BH}>10^9 M_\odot$ black hole is a factor $40000/8770$ greater than estimated 
previously (increasing the corresponding comoving density to $1.2\times 10^{-2}$ Gpc$^{-3}$).
In turn, the density of all radio--loud sources hosting a $M_{\rm BH}>10^9 M_\odot$ black hole
must be a factor $2\Gamma^2$ larger, namely 
$\sim 4.7(\Gamma/14)^2$ Gpc$^{-3}$ in the $5<z<6$ bin.
Note that this is a lower limit, since other high redshift blazars could exist, 
but not yet identified as such.

Moreover, assuming that the ratio of radio--loud to radio--quiet AGN is $\sim 0.1$,
the SMBHs with $M_{\rm BH}>10^9M_\odot$ have to be {\it at least} $10^4$ just after 
$10^9$ yrs from the birth of the Universe.

\section*{Acknowledgements}
We would like to thank the anonymous referee for useful
comments.
This publication makes use of data products from the Wide--field
Infrared Survey Explorer, which is a joint project of the University
of California, Los Angeles, and the Jet Propulsion
Laboratory/California Institute of Technology, funded by the National
Aeronautics and Space Administration.
Part of this work is based on archival data, software or on--line services 
provided by the ASI Data Center (ASDC).
Part of the funding for GROND
(both hardware as well as personnel) was generously granted from the Leibniz
Prize to Prof. G. Hasinger (DFG grant HA 1850/28--1).
This work made use of data supplied by the UK Swift
Science Data Centre at the University of Leicester

% \bsp

\label{lastpage}
\end{document}